# Electrochemical Storage Mechanism in oxy-Hydroxyfluorinated Anatase for Sodium-ion Batteries

Wei Li,[a] Mika Fukunishi,[b] Benjamin. J. Morgan,[c] Olaf. J. Borkiewicz,[d] Valérie Pralong,[e] Antoine Maignan,[e] Henri Groult,[a] Shinichi Komaba,[b] and Damien Dambournet*[a,f]

Replacing lithium ions as the charge carriers in rechargeable batteries with sodium ions can induce noticeable differences in the electrochemical storage mechanisms of electrode materials. Many material parameters, such as particle size, morphology, and the presence of defects are known to further affect the storage mechanism. Here, we report an investigation of how the introduction of titanium vacancies into anatase $TiO_2$ affects the sodium storage-mechanism. From pair distribution function analysis, we observe that sodium ions are inserted in titanium vacancies at the early stage of the discharge process. This is supported by density functional theory calculations, which predict that sodium insertion is more favourable at vacancies than at interstitial sites. Our calculations also show the intercalation voltage is sensitive to the anion coordination environment of the. Sodiation to higher concentrations induces a phase transition toward a disordered rhombohedral structure, similar to that observed in defect-free $TiO_2$. Finally, we find the x-ray diffraction pattern of the rhombohedral phase drastically changes depending on the composition and degree of disorder, providing further comprehension on the sodium storage mechanism of anatase.

## 1. Introduction

The search for new electrode materials for rechargeable sodium-ion batteries has been motivated by the large abundance of sodium in the Earth's crust, in contrast to the relatively limited amount of lithium, and the possibility of low-cost electrode processing, enabled by the replacement of copper with aluminium as the anodic current collector.[1,2] Moving from lithium to sodium insertion/intercalation chemistry is, however, not trivial, due to the potential for changes in storage mechanisms for lithium versus sodium, due to these ions different ionic radii and polarizabilities.[3,4]

The systematic understanding of sodium storage mechanism is a prerequisite to commercial use of sodium-ion electrode materials. Anatase $TiO_2$ is a fascinating material with multiple applications and provides a versatile platform that bridges fundamental and applied studies. Anatase $TiO_2$ displays attractive electrochemical properties with respect to working voltage (0.8 V vs. $Na^+$/Na), cycling stability, and high-rate performance.[5–8] A full understanding the influence of material parameters, such as size, morphology, and defect stoichiometry, on the sodium storage mechanism of anatase is, however, still lacking.[9–12]

Previous studies on the lithium insertion mechanism in anatase $TiO_2$ have shown that particle size and defect chemistry can have strong effects, with the possible suppression of the phase transition from the pristine tetragonal to the orthorhombic lithiated structure.[13–17] Advances in identifying sodium storage mechanism in anatase $TiO_2$ have revealed some apparent discrepancies, as highlighted by Yan et al.[18] A general observation was the loss of long-range order induced by sodium insertion in anatase.[11,19] To address these discrepancies, we have previously considered the origin of this disorder using high-energy total scattering data, combined pair distribution function (PDF) analysis and density functional theory (DFT) calculations: sodium insertion into anatase $TiO_2$ induces a symmetry change from the pristine tetragonal (space group I41/amd) to a layered-like rhombohedral symmetry (R-3m).[10] Because of the initial three-dimensionality symmetry of the anatase network, the stabilized layered $Na_xTiO_2$ structure shows high cationic intermixing between the sodium and titanium slabs.

To better understand how the presence of crystal defects affects this sodium storage mechanism, we have performed a combined PDF and DFT study of sodium insertion into anatase $TiO_2$ containing large concentrations of titanium vacancies (□). Comparing our results with those for nominally defect-free $TiO_2$ reveals that sodium ions are first inserted into titanium vacancies without any changes to crystal symmetry. Further sodium insertion, however, drives a transition to rhombohedral symmetry with similar reversible capacity as for stoichiometric $TiO_2$.

## 2. Experimental

### 2.1. Material synthesis and characterizations

Titanium oxy-hydroxyfluoride was prepared following the protocol described in our previous work.[17] A solution containing titanium isopropoxide (4 mL), aqueous HF (1.2 mL) and 27 mL of isopropanol was placed inside a Teflon-line container, sealed in an autoclave and further subjected to heat treatment at 90 °C for 12 hours. After washing, the powder was recovered and vacuum dried at 150 °C overnight. The chemical composition was $Ti_{0.78}\square_{0.22}O_{1.12}F_{0.40}(OH)_{0.48}$. Laboratory powder X-ray diffraction was carried out using a Rigaku Ultima IV X-ray diffractometer equipped with a Cu Kα radiation source (λ= 1.54059 Å). Chemical sodiation was performed using sodium naphtalenide as reducing agent. Anatase F-$TiO_2$ nanoparticles (200 mg) were dispersed in 20 mL of THF. Naphtalene (0.5 g) and small pieces of Na metal (170 mg) were added to the mixture and stirred in a glove box for 6 days at room temperature. Finally, the reduced phase was washed with THF and dried under vacuum.

*Magnetic properties.* The magnetic susceptibility, χ, as a function of temperature was obtained from magnetic moment measurements performed by squid magnetometry, with a MPMS-5T from Quantum Design. The data were collected upon warming from 5K to 300K in a 100 Oe applied magnetic field in zero-field-cooling (zfc) and field-cooling (fc) processes. A diamagnetic correction was applied to $\chi_{mol}$. Additional isothermal magnetic moment measurements were made as a function of the applied magnetic field H. To avoid air exposure, the sample was prepared in a glove box. About 0.1 g of the sample powder was set within a gelatine capsule.

**2.2. Electrochemistry**

Electrodes were prepared by mixing 80 wt % active material, 10 wt % acetylene black as the conductive agent, and 10 wt % sodium carboxymethyl cellulose[20] (CMC) as the binder. The mixture was dispersed in a diluted $H_2O_2$ aqueous solution ($H_2O_2$:$H_2O$ = 2.5:97.5 v/v %). The homogenous slurry was coated on an aluminum foil, then dried at 80 °C under air for 6 h and under primary vacuum for 12 h to evaporate the solvent. Sodium half cells were assembled inside a glove box filled with Ar. Sodium foil and a sandwich-like film composed of a glass filter in the two sides and polyolefin in the middle were used as the counter electrode and separator, respectively. A solution of 1 M $NaPF_6$ dissolved in ethylene carbonate and diethyl carbonate (EC:DEC) was used as the electrolyte. The sodium half cells were cycled in a voltage window 0.0-2.0 V vs. $Na^+$/Na at a current density 25 mA g$^{-1}$ using galvanostatic discharge-charge testers (TOSCAT-3100, Toyo System Co. Ltd.). Cycled cells were opened in the glovebox and washed with EC:DEC and DEC. Material was removed from the current collector and stored in the glove box for further characterization.

**2.3. Synchrotron X-ray diffraction**

Electrodes were loaded inside Kapton capillaries. Data were collected at the 11-ID-B beamline at the Advanced Photon Source at Argonne National Laboratory, using high energy X-rays (λ = 0.2128 Å) allowing access to high values of momentum transfer $Q_{max}$ ≈ 22 Å$^{-1}$.[21,22] The diffraction images were integrated within fit2D to obtain the one-dimensional diffraction data.[23] The G(r) function was extracted from the data using PDFgetX2,[24] after correcting for background and Compton scattering. The refinement of the PDF data was carried out using the PDFgui software.[25]

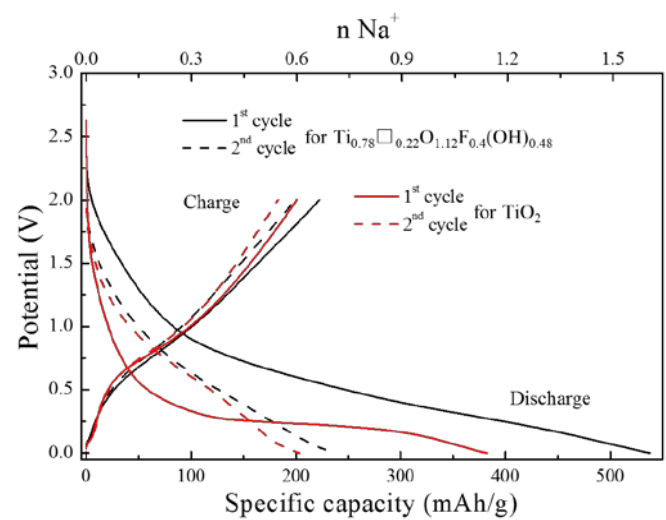

**Figure 1.** Galvanostatic discharge-charge curves of anatase $TiO_2$ and $Ti_{0.78}\square_{0.22}O_{1.12}F_{0.40}(OH)_{0.48}$ electrodes vs. sodium. The cells was cycled under 25 mA.g$^{-1}$.

**2.4. Computational**

The density functional theory (DFT) calculations were performed using the code VASP[26,27] with valence electrons described by a plane-wave basis, with a cutoff of 500 eV. Interactions between core and valence electrons were described using the projector augmented wave (PAW) method[28], with cores of [Ar] for Ti, [He] for O, [He] for F, [Ne] for Na, and [H$^+$] for H. All calculations used the revised Perdew-Burke-Ernzerhof generalized gradient approximation function PBEsol[29], with a Dudarev +$U$ correction applied to the Ti d states (GGA+$U$)[30,31]. We used a value of $U_{Ti,d}$=4.2 eV, which has previously been used to model intercalation of lithium and other metal ions in anatase $TiO_2$.[32–34] To model anatase $TiO_2$, we first performed a full geometry optimisation on a single $Ti_4O_8$ unit cell. Optimized lattice parameters were obtained by fitting a series of constant volume calculations to the Murnaghan equation of state. All subsequent calculations were fixed to these optimised lattice parameters, and used 3 × 3 × 1 supercells (108 atoms). Intercalation into stoichiometric anatase $TiO_2$ was modelled using with a single Na ion inserted at an interstitial site ($NaTi_{36}O_{72}$). Titanium vacancies were modelled as a defect complex comprising a Ti vacancy, with four charge-compensating anions $X$={OH,F} at adjacent equatorial sites. Na intercalation energies were calculated for 4$X$=(4F, 3F+OH, 2F+2OH, F+3OH, and 4OH). In the case of 4$X$=2F+2OH, we considered like anions arranged in adjacent (cis) and opposite (trans) equatorial site pairs. Individual calculations were deemed optimised when all atomic forces were smaller than 0.01 eV Å$^{-1}$. All calculations were spin polarized, and used a 4 × 4 × 2 Monkhorst-Pack grid for sampling k-space in the single unit cell, and a 2 × 2 × 2 grid for the 3 × 3 × 1 cells. To calculate intercalation energies, a reference calculations for metallic Na was performed using the same convergence criteria as above. We considered a 2-atom cell for Na, with a 16 × 16 × 16 Monkhorst-Pack grid for k-space sampling. A data set containing all DFT calculation inputs and outputs is available at the University of Bath Data Archive[35], published under the CC-BY-SA-4.0 license. Analysis scripts containing intercalation energy calculations, and code to produce Figure S4 are available as an open-source repository as reference[36], published under the MIT license.

## 3. Results and discussion

### 3.1. Electrochemical properties

The impact of the titanium vacancies on the sodium storage mechanism of anatase was first assessed using galvanostatic discharge-charge curves. **Figure 1** gathers the galvanostatic data obtained for the first two cycles for anatase $TiO_2$ and $Ti_{0.78}\square_{0.22}O_{1.12}F_{0.40}(OH)_{0.48}$ vs. $Na^+$/Na.

During the first discharge, the voltage of the cell gradually decreases upon sodiation reaching a capacity of 538 mAh.g$^{-1}$. The smooth discharge curve for $Ti_{0.78}\square_{0.22}O_{1.12}F_{0.40}(OH)_{0.48}$ electrode contrasts with the observation of an irreversible plateau at 0.3 V for $TiO_2$. During the first charge, the capacity reached 222 mAh.g$^{-1}$ with a sloping aspect similar to those of pure $TiO_2$ suggesting a similar storage mechanism. This is confirmed with the subsequent potential-capacity curves that showed similar features for both anatase TiO$_2$ and Ti$_{0.78}\square_{0.22}$O$_{1.12}$F$_{0.40}$(OH)$_{0.48}$. Upon prolonged cycling, both anatase showed similar capacities **(Supporting Info)** of around 150 mAh.g$^{-1}$ after 50 cycles.

### 3.2. Structural analysis

The signature of the first discharge curve for anatase TiO$_2$ and Ti$_{0.78}\square_{0.22}$O$_{1.12}$F$_{0.40}$(OH)$_{0.48}$ pointed out differences in reactivity particularly during the early stage of the sodiation. In our previous study, we showed that the sodiation of TiO$_2$ up to 100 mAh.g$^{-1}$ and characterized by a rapid drop of the voltage (0.3 V), did not induce any structural change and was due to electrolyte decomposition.[10] In the case of Ti$_{0.78}\square_{0.22}$O$_{1.12}$F$_{0.40}$(OH)$_{0.48}$ electrode, we observed a smooth decrease of the voltage. To compare the early stage of the sodiation, we discharged an Ti$_{0.78}\square_{0.22}$O$_{1.12}$F$_{0.40}$(OH)$_{0.48}$ electrode up 100 mAh.g$^{-1}$ and measured high-energy total scattering data from which we obtained the pair distribution function, denoted G(*r*). We applied a real-space refinement to extract structural information. The best fit to the PDF data was obtained by using a single phase having a tetragonal symmetry similar to the pristine electrode. We hypothesized that sodium can be inserted in the titanium vacancies and accordingly refined the rate occupancy of Na in the Ti (4a) site. Moreover, we refined the rate occupancy of the interstitial site (4b Wyckoff site). **Figure 2** shows the final PDF refinement within the inter-atomic distances, *r*, from 1 to 30 Å. The structural parameters obtained after the refinement was gathered in **Table 1**.

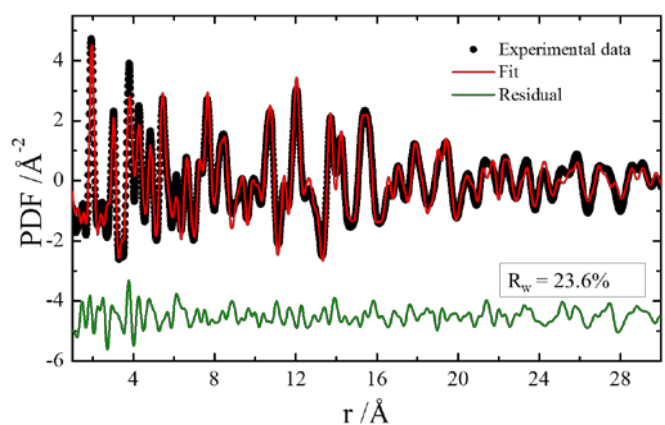

**Figure 2.** PDF refinement of Ti$_{0.78}\square_{0.22}$O$_{1.12}$F$_{0.40}$(OH)$_{0.48}$ electrode discharged to 100 mAh.g$^{-1}$ (~0.85 V, first discharge).

The refinement of the rate occupancy confirmed that sodiation occurred selectively within the titanium vacancies (up to 0.17 Na+) rather than in the interstitial site. The calculated capacity based on the number of sodium obtained by PDF refinement yielded 75 mAh.g$^{-1}$ out of the 100 mAh.g$^{-1}$ measured suggesting a capacity contribution due to the electrolyte decomposition.

**Table 1.** Structural parameters of pristine Ti$_{0.78}\square_{0.22}$O$_{1.12}$F$_{0.40}$(OH)$_{0.48}$ and discharged electrode to 100 mAh.g$^{-1}$ capacity.

| Parameters | Pristine | Discharged (100 mAh.g$^{-1}$) |
|---|---|---|
| a [Å] | 3.792 (1) | 3.805 (2) |
| c [Å] | 9.481 (5) | 9.490 (6) |
| V [Å$^3$] | 136.3 | 137.4 |
| **Rate occupancy** | | |
| Ti (4a) | 0.78(3) | 0.76(2) |
| Na (4a) | N.A | 0.17(3) |
| Na (4b) | N.A | 0.04(2) |

To better understand the energetics of sodium intercalation at vacancies, we performed DFT calculations of sodium insertion at a single vacancy site. To do so, we first considered a supercell containing one vacancy formed by fluorine doping, i.e., Ti$_{35}\square_1$O$_{68}$F$_4$. For this F-compensated vacancy, we calculated a Na-intercalation energy of -1.88 eV. This is much more favourable than insertion at an interstitial site, which we calculate to be +0.03 eV. Experimentally, the insertion of sodium within the vacancies occurs in a large potential window ranging from around 2 to 1 V. To understand such a large potential window, we considered the effect of the anionic environment of the vacancy, particularly the presence of both fluorine and OH groups. Further calculations were performed with equivalent supercells where the group of four charge-compensating anions adjacent to the vacancy were varied from a fully fluorinated to fully hydroxylated environment. The incorporation of OH groups causes the intercalation energy to increase from -1.88 eV (4F) to -0.22 eV (4OH) (SI, Figure S4), corresponding to a decrease in insertion voltage. These results show a pronounced effect of the vacancy anionic environment on the voltage reaction and account for the observed large potential window observed.

After full discharge, high-energy X-ray diffraction showed that Bragg peaks related to anatase completely disappeared while new broad Bragg peaks appear with pattern similar to the O3-type (R-3m) Na$_x$TiO$_2$ phase (SI, Figure S2). We recall that the "O3" nomenclature referred to Delmas' notation indicating that the Na ions are octahedrally coordinated by oxygen and display a repetition period of three transition metal stacking along the c-axis.[37]

The observation of sodium-induced structural changes from three-dimensional anatase to layered-like Na$_x$TiO$_2$ is similar to that observed in defect-free TiO$_2$.[10] The structure of the fully sodiated anatase Ti$_{0.78}\square_{0.22}$O$_{1.12}$F$_{0.40}$(OH)$_{0.48}$ was examined by PDF refinement. The structural refinement was performed according to our previous published procedure.[10] Accordingly, the analysis comprises refining the rate occupancy of the 3a and 3b cation sites and the intermixing of Na and Ti atoms in the Na/Ti slabs. The final refinement is shown in **Figure 3**. **Table 2** gathers the structural parameters obtained after the refinement and includes those of ordered[38] and disordered[10] Na$_x$TiO$_2$ and Na$_{0.54}$TiO$_2$[39] compounds.

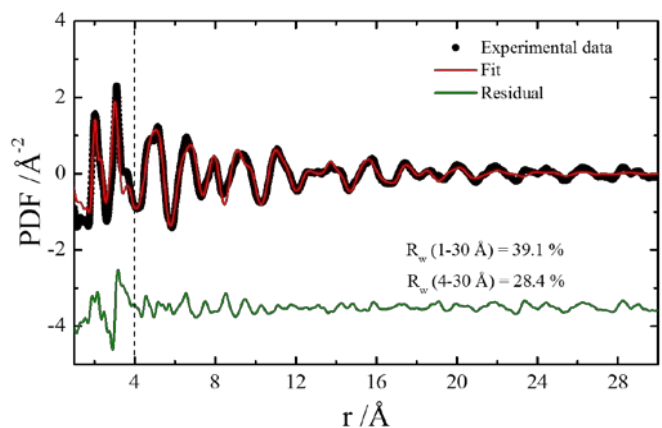

**Figure 3.** PDF refinement of fully sodiated Ti$_{0.78}$□$_{0.22}$O$_{1.12}$F$_{0.40}$(OH)$_{0.48}$ electrode (~0 V, first discharge).

A good fit was obtained for the *r* region beyond 4 Å with a reliability factor of $R_w$ = 28.4% (4 – 30 Å) while extension to the 1-30 Å region cannot captured the local structure ($R_w$ = 39.1 %). The observation of local structural disorder was preliminary assigned to the cation mixing in the octahedral Na and Ti sites as shown by the refinement of the rate occupancy. Such a cation mixing yield to a contraction along the *c*-axis because of the presence of high valence Ti$^{IV}$/Ti$^{III}$ ions in the Na layers.[40]

To summarize on the discharge process, the presence of titanium vacancies allow accommodating Na$^+$ ions at the early stage of the sodiation. Higher insertion of Na$^+$ ions results in a symmetry change to the rhombohedral structure as observed in defect-free anatase. The presence of titanium vacancies thus does not modify the electrochemical storage mechanism as found in the lithium case which was assigned to the suppression of the edge-shared pairs of LiO$_6$ octahedra.[17,33]

**Table 2.** Structural parameters of fully discharged pristine Ti$_{0.78}$□$_{0.22}$O$_{1.12}$F$_{0.40}$(OH)$_{0.48}$, ordered and disordered Na$_x$TiO$_2$ and Na$_{0.54}$TiO$_2$ compounds.

| Parameters | Fully discharged (this work) | Disordered Na$_x$TiO$_2$ (ref [10]) | Ordered Na$_x$TiO$_2$ (ref [38]) | Na$_{0.54}$TiO$_2$ (ref [39]) |
|---|---|---|---|---|
| a [Å] | 3.010(9) | 2.989(8) | 3.05414(4) | 2.9791(6) |
| c [Å] | 15.36(9) | 15.07(9) | 16.2412(2) | 16.9280(3) |
| V [Å$^3$] | 120.53 | 116.60 | 131.20 | 130.11 |
| Rate occupancy | | | | |
| Na(3a) | 0.60(20) | 0.43(25) | 0.984(7) | 0.54(1) |
| Ti (3a) | 0.38(12) | 0.57(25) | 0.016(7) | 0.0 |
| Na (3b) | 0.16(18) | 0.39(14) | 0.016(7) | 0.0 |
| Ti (3b) | 0.54(13) | 0.39(12) | 0.984(7) | 1.0 |

During the charge process, we observed a continuous loss of long-range order with PDF features characteristic of anatase local structural type **(SI, Figure S2)**.

### 3.3. Simulated XRD patterns of the Na$_x$TiO$_2$ system

In the Na$_x$TiO$_2$ system, both the composition (x value) and the cationic slabs ordering impact the structure and hence the X-ray diffraction pattern. The variation of the Bragg peak position and intensities can yield to erroneous interpretation of the XRD data and possibly to wrong mechanism. To exemplify how the X-ray diffraction pattern can vary in the Na$_x$TiO$_2$ system, we simulated the XRD data (Powder Cell[41]) of the compounds presented in the Table 2 **(Figure 4)**. The large variation of the *c*-parameter value induces a broad set of Bragg peak positions for the 00l lines. For instance, the position of the (003) Bragg peak varies from 15.7° for Na$_{0.54}$TiO$_2$ to 17.3° (2θ) for sodiated Ti$_{0.78}$□$_{0.22}$O$_{1.12}$F$_{0.40}$(OH)$_{0.48}$. In addition, the (003) peak intensity indicates the degree of layered ordering between Na and Ti ions. The fully ordered Na$_{0.54}$TiO$_2$ shows the maximum intensity while introducing only 2% of Na/Ti exchange yields to a decrease of 57 % of the peak intensity. For the sodiated TiO$_2$ electrode, the high Na/Ti exchange leads to the peak extinction pointing high disorder. The resulting XRD pattern shows features of a cubic symmetry which can lead to misinterpretation in the storage mechanism.[42]

Note that the degree of layered ordering between Na and Ti ions in the R-3m structure impacts its properties. Wu et al[43] reported that the layered Na$_{0.99}$TiO$_2$ undergoes a reversible O3 to O'3 phase transition during sodiation/de-sodiation mechanism.

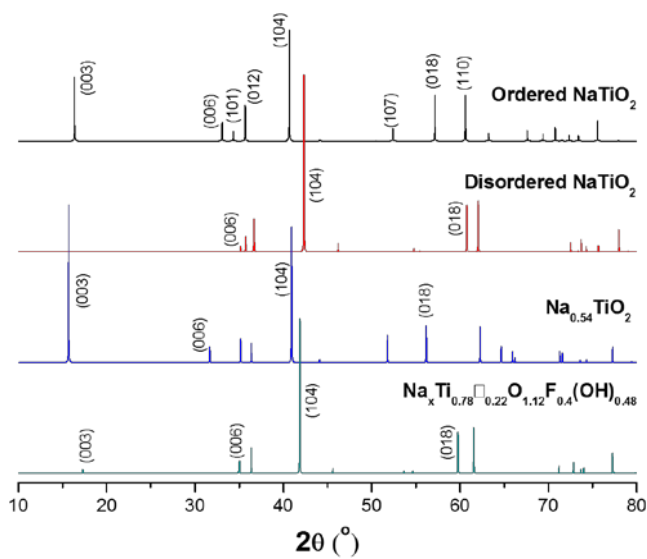

**Figure 4.** Simulated XRD patterns of Na$_x$TiO$_2$ compounds.

Structural features can also largely affect magnetic properties. To assess of the magnetic properties of the reduced phase, we performed a chemical reduction. The XRD pattern of the chemically reduced phase confirmed the phase transition toward the rhombohedral structure **(SI, Figure S3)**. The magnetic susceptibility (χ), corrected from the diamagnetism of ions, measured as a function of the temperature for the chemically sodiated sample is shown in **Figure 5**. The obtained χ values are very small and rather T independent from 50K to 300K. An attempt to fit the reciprocal magnetic susceptibility by a Curie-Weiss law leads to a very low μ$_{eff}$ value of 0.69 μB which is not compatible with a paramagnetism coming from the spin S=1/2 of Ti$^{3+}$ (1.73μ$_B$ for 3d$^1$). It indicates rather a Pauli paramagnetism as in the Na$_x$TiO$_2$ anatase.[10] As expected, the

S=0 $Ti^{4+}$ oxidized sample shows also a T independent χ (inset of Fig. 6) which χ >0 values come from the sample holder. The lack of magnetic ordering detected on the χ(T) curve of the sodiated sample could be explained by electronic conduction with delocalized charges in this 2D structure (the powder is black) and by both nanometer size of the grains and presence of Ti vacancy which are also not favorable for a S=1/2 long range magnetic moment ordering.

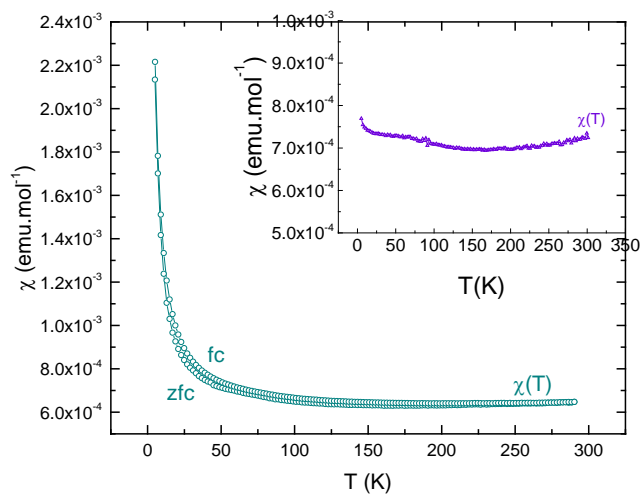

**Figure 5**. Susceptibility versus temperature for $Na_xTi_{0.78}\square_{0.22}O_{1.12}F_{0.40}(OH)_{0.48}$ sample. Inset: Susceptibility versus temperature for the as-prepared $Ti_{0.78}\square_{0.22}O_{1.12}F_{0.40}(OH)_{0.48}$ sample.

## 4. Conclusions

Here, we intended to investigate the effect of the presence of titanium vacancies on the sodium storage mechanism of anatase. By comparing the electrochemistry of anatase $TiO_2$ and vacancy containing structure $Ti_{0.78}\square_{0.22}O_{1.12}F_{0.40}(OH)_{0.48}$, we showed differences on the galvanostatic discharge curves. Notably, we showed that at the early stage of sodiation, $Ti_{0.78}\square_{0.22}O_{1.12}F_{0.40}(OH)_{0.48}$ electrode presents a smooth variation of the voltage vs. capacity which contracts with the rapid drop of the voltage observed for defect-free $TiO_2$. A structural analysis performed by means of pair distribution function obtained on sodiated electrode demonstrated that sodium ions are inserted into titanium vacancies. DFT-calculations support the favourable sodium insertion in the titanium vacancies compared to interstitial sites. Further calculations demonstrate that the nature of the local anionic environment of the vacancy strongly impacts the expected insertion voltage. Similar to stoichiometric $TiO_2$, the full sodiation of $Ti_{0.78}\square_{0.22}O_{1.12}F_{0.40}(OH)_{0.48}$ induces a structural change from the tetragonal to an O3 rhombohedral type structure built from Ti and Na slabs featuring high cationic disorder. We further showed that the Bragg peaks position and intensity of the x-ray diffraction pattern of the rhombohedral phase is largely influence by the composition and disorder of the structure which can lead to misinterpretation of the data. Finally, we showed that the subsequent charge/discharge processes are similar between anatase $TiO_2$ and vacancy containing structure $Ti_{0.78}\square_{0.22}O_{1.12}F_{0.40}(OH)_{0.48}$.

## Acknowledgements

This research used resources of the Advanced Photon Source, a U.S. Department of Energy (DOE) Office of Science User Facility operated for the DOE Office of Science by Argonne National Laboratory under Contract No. DE-AC02-06CH11357. B. J. M. acknowledges support from the Royal Society (UF130329). DFT calculations were performed using the Balena High Performance Computing Service at the University of Bath, and using the ARCHER supercomputer, with access through membership of the UK's HPC Materials Chemistry Consortium, funded by EPSRC grant EP/L000202.

**Data Access Statement**. The DFT dataset supporting this study is available from the University of Bath Research Data Archive (doi:10.15125/BATH-00473),[36] published under the CC-BY-SA-40 license. This dataset contains all input parameters and output files for the VASP DFT calculations, and Python scripts for collating the relevant data used in our analysis. A Jupyter notebook containing code to produce Figure S4 is also available (Ref 36, doi: 10.5281/zenodo.1181872), published under the MIT license.

# Electrochemical Storage Mechanism in oxy-Hydroxyfluorinated Anatase for Sodium-ion Batteries

Wei Li,[a] Mika Fukunishi,[b] Benjamin. J. Morgan,[c] Olaf. J. Borkiewicz,[d] Valérie Pralong,[e] Antoine Maignan,[e] Henri Groult,[a] Shinichi Komaba,[b] and Damien Dambournet*[a,f]

a. Sorbonne Université, CNRS, Physico-chimie des électrolytes et nano-systèmes interfaciaux, PHENIX, F-75005 Paris, France. Email: damien.dambournet@sorbonne-universite.fr
b. Department of Applied Chemistry, Tokyo University of Science, 1-3 Kagurazaka, Shinjuku, Tokyo 162-8601, Japan
c. Department of Chemistry, University of Bath, Claverton Down, BA2 7AY, United Kingdom
d. X-ray Science Division, Advanced Photon Source, Argonne National Laboratory, 9700 South Cass Avenue, Argonne, Illinois 60439, United States
e. Laboratoire CRISMAT, ENSICAEN, Université de Caen, CNRS, 6 Bd Maréchal Juin, F-14050 Caen, France
f. Réseau sur le Stockage Electrochimique de l'Energie (RS2E), FR CNRS 3459, 80039 Amiens, France

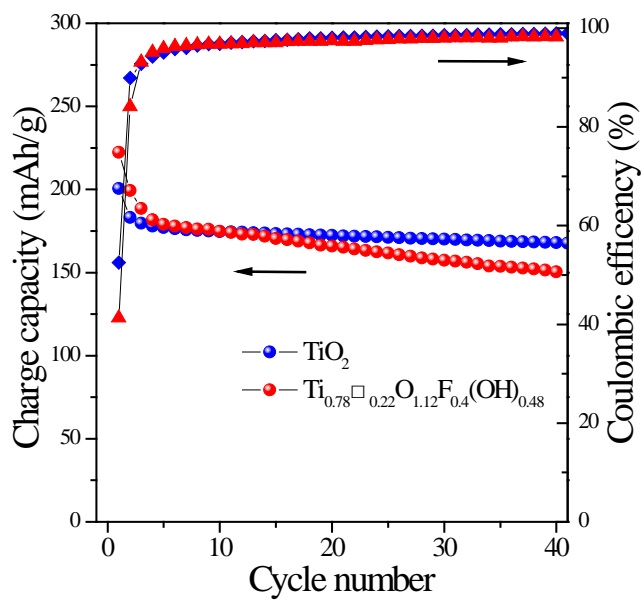

**Figure S1.** Cycling behavior of $Ti_{0.78}\square_{0.22}O_{1.12}F_{0.4}(OH)_{0.48}$ *vs.* $TiO_2$. The cell was cycled between 0 – 2 V at a current density of 25 mA/g.

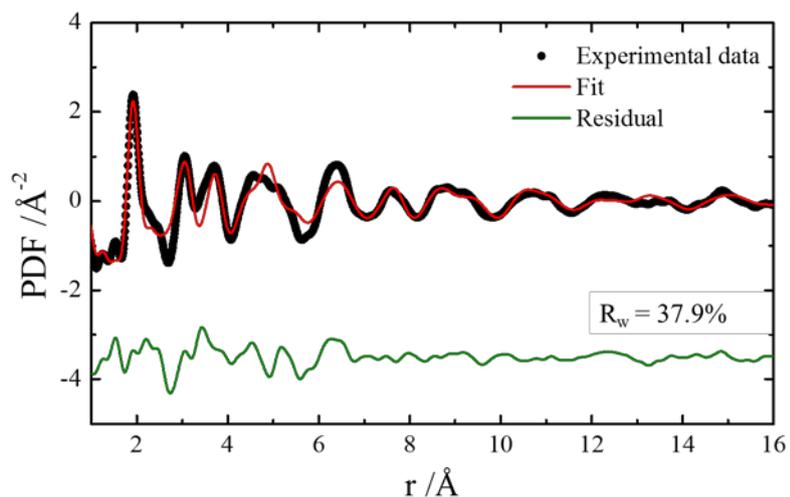

**Figure S2.** PDF refinement of the electrode charged to 2.0 V using TiO$_2$ anatase and O3-type Na$_x$TiO$_2$ (space group: R-3m) model.

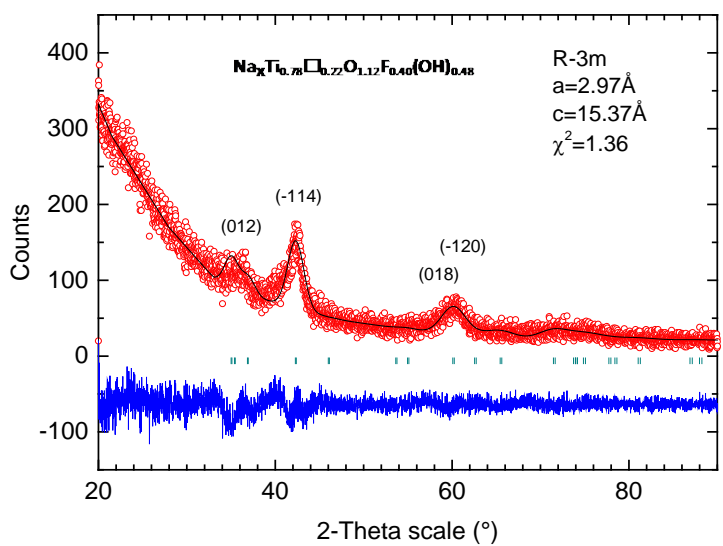

**Figure S3.** Le Bail profile refinement of the x-ray diffraction powder pattern of the reduced Na$_x$Ti$_{0.78}$□$_{0.22}$O$_{1.12}$F$_{0.40}$(OH)$_{0.48}$ obtained by chemical sodiation.

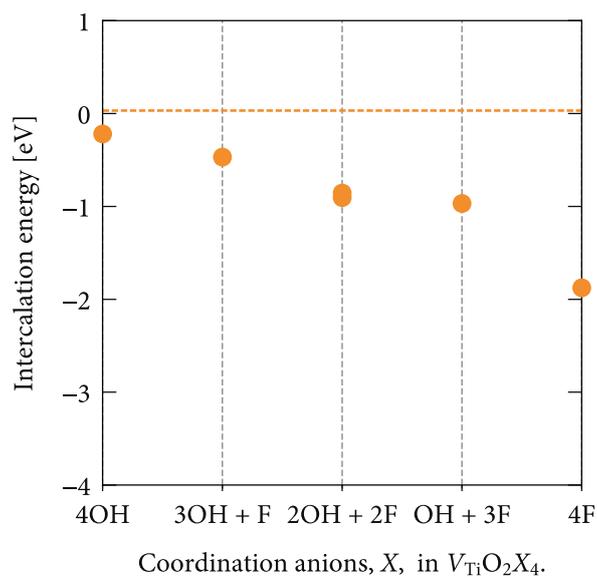

**Figure S4.** Intercalation energy of lithium in a $Ti_{35}O_{68}X_4$ supercell with $X$ = F⁻, OH⁻. The horizontal dashed line shows the intercalation energy for lithium in stoichiometric anatase $TiO_2$. Source: The data set and code to generate this figure, and the figure file, are available under the MIT licence as part of [1].